\documentstyle[12pt,epsfig]{article}
\begin{document}
\baselineskip=22pt
 ~~~~~~~~~~~~~~~~~~~~~~~~~~~~~~~~~~~~~~~~~~~~~~~~~~~~~~~~~~~~~~~~~~~~~~~~~~BIHEP-TH-2002-56

\vspace{1.2cm}

\begin{center}
{\Large \bf Minkowski Brane in Asymptotic dS$_5$ Spacetime without
Fine-tuning }

\bigskip

Zhe Chang\footnote{changz@mail.ihep.ac.cn}, ~Shao-Xia Chen\footnote{ruxanna@mail.ihep.ac.cn}~and~Xin-Bing Huang\footnote{huangxb@mail.ihep.ac.cn}\\
{\em Institute of High Energy Physics\\
Chinese Academy of Sciences \\
P.O.Box 918(4), 100039 Beijing, China}
\end{center}

\vspace{1.2cm}
\begin{abstract}
We discuss properties of a $3$-brane in an asymptotic
5-dimensional de-Sitter spacetime. It is found that a Minkowski
solution can be obtained without fine-tuning. In the model, the
tiny observed positive cosmological constant is interpreted as a
curvature of 5-dimensional manifold, but the Minkowski spacetime,
where we live, is a natural $3$-brane perpendicular to the fifth
coordinate axis.

\end{abstract}
\vspace{1.2cm} PACS numbers: 98.80.-k, 11.25.-w, 04.50.+h, 98. 80.
Es \vspace{1.2cm}

The tiny but not zero cosmological constant has puzzled
astrophysicists and physicists for nearly a
century\cite{weinberg}. Various attempts have been made in trying
to solve the puzzle. However, up to now, there hasn't been a
theory that can give a cosmological constant whose order is the
same as that of the observed value.

New progress made in this direction came from the brane world
picture of cosmology, and in particular, the Randall-Sundrum (RS)
scenario\cite{RS1, RS2}. The RS model was proposed by Randall and
Sundrum to aim at a reasonable explanation of the hierarchy
between the electro-weak scale and Planck scale in the
four-dimensional effective field theory. In this scenario, the
spacetime is five-dimensional. We live in a 3-brane, which is
perpendicular to the fifth coordinate axis. The fifth dimension
can be compact or noncompact. All matter and interactions except
gravity are confined to the 3-brane. Since the RS model was
proposed, many physicists\cite{cp3}-\cite{cp7} have been trying to
apply it to such issues as the hierarchy problem\cite{ch1},
cosmological constant
problem\cite{cp3,cp5,cp6,ch1,cp1,cp4,cp2,cp7}, localization of
gravity\cite{lg1}, dynamics of the brane\cite{db1}, and the
realization of the $AdS_5$ spacetime in string theory.

In this Letter,  we consider an asymptotic $dS_5$ spacetime, which
is characterized by a positive cosmological constant. It is found
that there is a natural solution of Minkowski brane  without
fine-tuning.

We start from the  action of the dS $5$-dimensional spacetime with
a dilaton field $\phi$ ($\equiv\phi(x_5)$)

\begin{equation}
\label{s}
 S=\int
d^5x\left(\sqrt{-G}\left(R-\frac{4}{3}(\nabla\phi)^2-\Lambda
e^{a\phi}\right)-\sqrt{-g}\delta(x_5)Ve^{b \phi}\right)~,
\end{equation}
where $G$ and $g$ are determinants of the five-dimensional
spacetime metric $G_{M N}$ and the four-dimensional brane metric
$g_{\mu\nu}$, respectively. The Einstein equation corresponding to
the action reads

\begin{equation}
\label{E}
\begin{array}{l}
\displaystyle\sqrt{-G}\left(R_{MN}-\frac{1}{2}G_{MN}R\right)-\frac{4}{3}\sqrt{-G}
\left(\nabla_M\phi\nabla_N\phi-\frac{1}{2}G_{MN}(\nabla\phi)^2\right)~~~~~~~~~~\\[0.5cm]
~~~~~~~~~~~~~~~~~~~~~~~~\displaystyle+\frac{1}{2}\left(\Lambda
e^{a\phi}\sqrt{-G}G_{MN}+\sqrt{-g}Ve^{b\phi}g_{\mu\nu}\delta^\mu_M\delta^\nu_N\delta(x_5)\right)=0~.
\end{array}
\end{equation}
The equation of motion of the dilaton is as following

\begin{equation}
\label{V} \sqrt{-G}\left(\frac{8}{3}\nabla^2 \phi-a\Lambda
e^{a\phi}\right)-b\sqrt{-g}V\delta(x_5)e^{b\phi}=0~.
\end{equation}
Following Randall and Sundrum, we assume that the metric of the
five-dimensional spacetime  is of the form

\begin{equation}
\label{G}
 ds^2=e^{2A(x_5)}g_{\mu\nu}dx^\mu dx^\nu+(dx^5)^2~,
\end{equation}
where
\begin{equation}
\label{g} g_{\mu\nu}={\rm
diag}\left(-e^{2\sqrt{-\bar{\Lambda}}x_4},~
e^{2\sqrt{-\bar{\Lambda}}x_4},~ e^{2\sqrt{-\bar{\Lambda}}x_4},~
1\right)~.
\end{equation}
Please note that, in general, the $3-$brane can be any symmetric
space.  With the above ansatz for the metrics on 5-dimensional
spacetime and $3$-brane, we transform the equations(\ref{V}) and
(\ref{E}) into the form

\begin{equation}
\label{d}
\frac{8}{3}\phi^{\prime\prime}+\frac{32}{3}A^\prime\phi^\prime-a\Lambda
e^{a\phi}-bV\delta(x_5)e^{b\phi}=0~,
\end{equation}
\begin{equation}
\label{44}
 3A^{\prime\prime}+\frac{4}{3}(\phi^\prime)^2+3\bar{\Lambda} e^{-2A}+\frac{1}{2}V
e^{b\phi}\delta(x_5)=0~,
\end{equation}
\begin{equation}
\label{55}
 6(A^\prime)^2-6\bar{\Lambda}
e^{-2A}-\frac{2}{3}(\phi^\prime)^2+\frac{1}{2}\Lambda
e^{a\phi}=0~,
\end{equation}
where the notation $^\prime$ denotes differentiation with respect
to $x_5$. We are interest on finding a natural Minkowski solution
of the brane system without fine-tuning.

Away from the $3$-brane, the equations of motion of the system
(bulk equations) reduce to the form

\begin{equation}
\label{db}
\frac{8}{3}\phi^{\prime\prime}+\frac{32}{3}A^\prime\phi^\prime-a\Lambda
e^{a\phi}=0~,
\end{equation}
\begin{equation}
\label{44b}
 3A^{\prime\prime}+\frac{4}{3}(\phi^\prime)^2+3\bar{\Lambda}
 e^{-2A}=0~,
\end{equation}
\begin{equation}
\label{55b}
 6(A^\prime)^2-6\bar{\Lambda}
e^{-2A}-\frac{2}{3}(\phi^\prime)^2+\frac{1}{2}\Lambda
e^{a\phi}=0~.
\end{equation}
The difficulty in trying to solve the bulk equations comes from
the $\bar{\Lambda}$ terms.  Here, we present a  natural Minkowskin
solution of the bulk equations of motion in integration form

\begin{equation}
\label{solution}
\begin{array}{rcl}
&&
A(x_5)=-\displaystyle\frac{2\alpha}{a}\ln{\left(a\sqrt{D}\left[1-2H(x_5)\right]x_5+2d\right)}
-e^{c-\left[2H(x_5)-1\right]x_5}+hx+\eta~,
\\ [0.5cm] &&
\displaystyle
\phi(x_5)=-\left[2H(x)-1\right]\displaystyle\int_{0^\pm}^{x_5}dx\sqrt{\frac{D}{\left(\frac{a\sqrt{D}\left[1-2H(x)
\right]x}{2}+d\right)^2}+\frac{9}{4}e^{c-\left[2H(x)-1\right]x}}~,\\
[0.5cm] && \bar{\Lambda}=0~,
\end{array}
\end{equation}
where $D\equiv\displaystyle\frac{3a\Lambda}{4(a+8\alpha)}~,$
$c,~h,~d,~\alpha$ and $\eta$ are constants, and $H(x)$ is the
Heviside function
$$
H(x)=\left\{ \begin{array}{cc} 1,~~~&~~{\rm for}~~x>0~,\\
0,~~~&~~{\rm for}~~x<0~.
\end{array}
\right.
$$
In figures 1 and 2, we give a plot of the dilaton field
$\phi(x_5)$ with selected parameters
$a_-=0.189,~d_-=1.545,~c_-=-28.481,~\Lambda_-=10^{-4},~\alpha=8.776
~ {\rm and}~ h_-=-4.01\times 10^{-3} $ for $x<0$;
$a_+=0.01,~d_+=15.875,~c_+=-19.63,~\Lambda_+=10^{-4},~\alpha=413.56
~ {\rm and}~ h_+=-8.306\times 10^{-3} $ for $x>0$.

\begin{center}
  \includegraphics{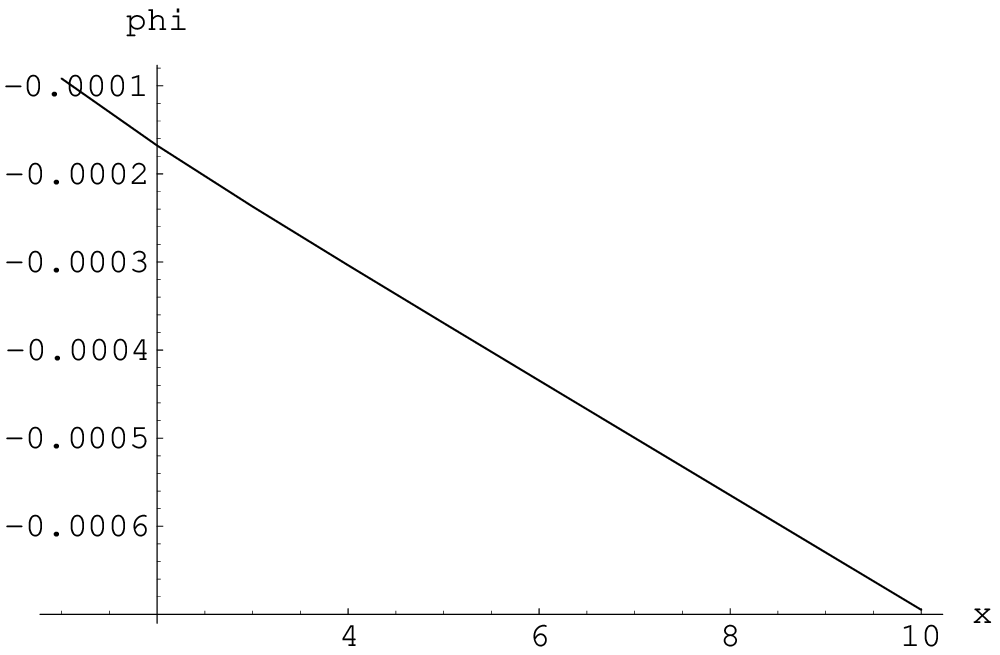}

{\bf FIG.1}~~~~ {  The behavior of the dilaton field $\phi(x)$.}
\end{center}
\begin{center}
  \includegraphics{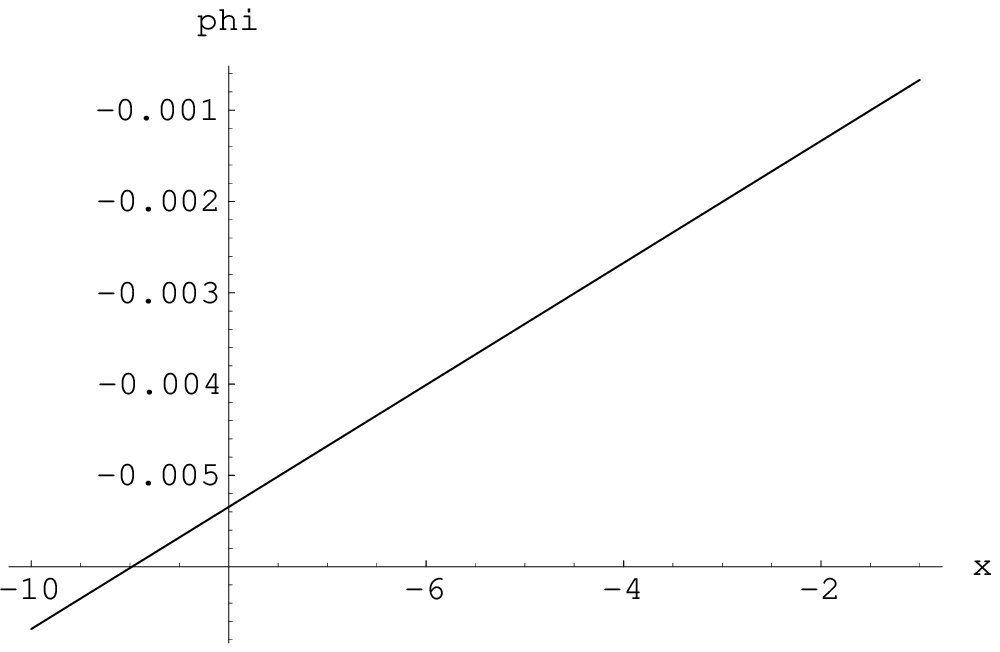}

{\bf FIG.2}~~~~{   The behavior of the dilaton field $\phi(x)$ for
$x<0$.}
\end{center}
 The matching conditions on the  brane for Minkowski solution
(\ref{solution}) are of the form

\begin{equation}
\label{self1}
\sqrt{\frac{D_+}{d_+^2}+\frac{9}{4}e^{c_+}}+\sqrt{\frac{D_-}{d_-^2}+\frac{9}{4}e^{c_-}}=-\frac{3}{8}bV
e^{b\phi(0)}~,
\end{equation}
\begin{equation}
\label{self2}
-\frac{8\sqrt{D_+}}{9a_+d_+}+e^{c_+}+h_+-\frac{8\sqrt{D_-}}{9a_-d_-}+e^{c_-}-h_-=-\frac{1}{6}V
e^{b\phi(0)}~.
\end{equation}
\vspace{0.4cm}

 {\bf Acknowledgements}: The work was supported in
part by the NSF of China.

\end{document}